\documentclass{iauJDSS}
\usepackage{graphics}

\title[Surviving on Mars?] 
{Surviving on Mars: test with LISA simulator}

\author[Galletta et al.]   
{G. Galletta$^{1,2}$,M. D'Alessandro$^3$, G. Bertoloni$^4$, F. Castellani$^4$, R. Visentin$^{1,4}$}

\affiliation{$^1$Dipartimento di Astronomia, Universit\`a di Padova, vicolo Osservatorio, 3 - I 35122 Padova, Italy  -  email: giuseppe.galletta@unipd.it\\[\affilskip]
$^2$CISAS "G. Colombo", Universit\`a di Padova 
$^3$ INAF, Osservatorio Astronomico di Padova 
$^4$ Dipartimento di Istologia, Microbiologia e Biotecnologie Mediche, Universit\`a di Padova
}

\pubyear{2009}
\volume{Volume 15}  
\pagerange{000--001}
\date{18 September 2009 and in revised form ??}
\jname{Highlights of Astronomy, Volume 15}
\editors{Ian F Corbett, ed.}
\begin{document}

\maketitle

\begin{abstract}
We present the results of some experiments performed in the Padua simulators of planetary environments, named LISA, used to study the limit of bacterial life on the planet Mars. The survival of {\it Bacillus} strains for some hours in Martian environment is shortly discussed. 

\end{abstract}

\firstsection 
\section{Introduction}

Planet Mars is a typical target for exploring the possibility that life arose in a planet different than Earth. Its present surface conditions are very inhospitables, if compared with most of the terrestrial environments, but several lifeforms have shown a strong capability to adapt to very harsh conditions and to survive even for some period in circumterrestrial space. If life, similar or different from the terrestrial one, existed on Mars in the past, some  life forms could have adapted to the climate changes and could have survived in some ecologic niche, near the soil or deep under the surface. Moreover, while exploring Mars, we may incidentally or voluntary drop terrestrial lifeforms, able to survive and to be reactivated once brought back in laboratories. Finding conditions that allow the survival of lifeforms in a Martian environment can have a double value: increasing the hope of finding extraterrestrial life and defining the limits for a terrestrial contamination of planet Mars.

\section{Simulating Mars}

In order to understand if some lifeforms may have survived on Mars, we conceived and built two simulators of its environments \cite{LISA2006, LISA2007} where to perform researches with bacteria strains: LISA (Laboratorio Italiano Simulazione Ambienti), that allows six simultaneous experiments, and a single-experiment version of it (mini-LISA). Our LISA environmental chambers may reproduce the conditions of many Martian locations near the surface (temperature ranging from 133K to 293K, atmospheric composition with the 95\% of CO2 at a pressure between 6 and 9 mb, strong UV radiation). Since we use a 500 liters reservoir of liquid nitrogen, refuelled once per week, experiments not longer than 25 hours may be performed inside LISA, while in mini-LISA there is theoretically no time limit. 

We proceeded keeping in mind two important caveats: First, we don't have (yet) neither Martian bacteria nor Martian soils to use for the tests, and so we must use terrestrial surrogates. Second, lifeforms on a different planet may have a fully different combination of nucleic acids and aminoacids, so our conclusions could be fully wrong if applied to a ``Martian life''. However, they are useful for contamination- or terraforming- studies.  

Inside our simulators we have studied the survival of several bacterial strains belonging to the genus {\it Deinococcus}, and to the endospore forming genera {\it Bacillus} and {\it Clostridium}. Cellular or endospores suspensions were layered on sterile coverslip dehydrated under sterile air flux, introduced in dedicated plates and then exposed to a pressure of 7.5 mbar of CO$_2$ and to $\sim$4 W m$^{-2}$ of UVC light. We simulated both ``Martian summer day'' (23 $^\circ$C) and ``winter day''(-80 $^\circ$C) conditions for several hours.

\section{Results and conclusions}

\begin{figure}
 \resizebox{13.5cm}{!}{\includegraphics{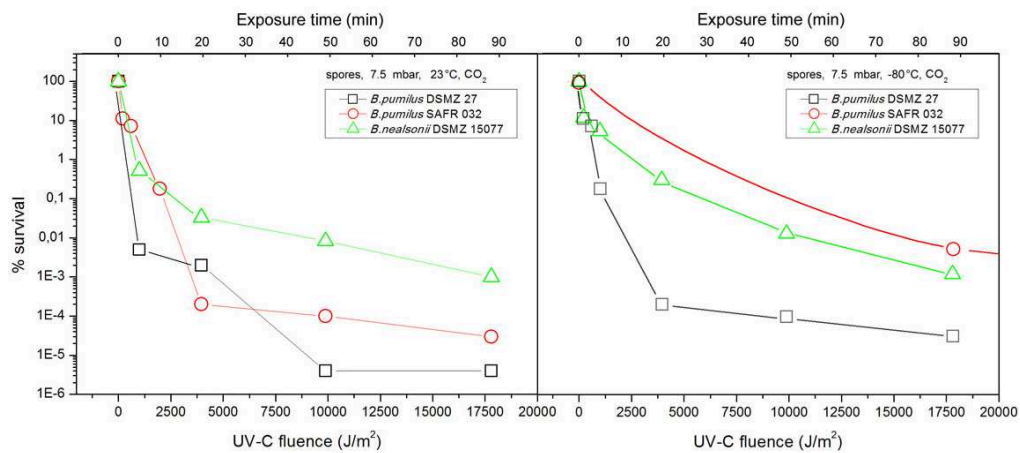}}
  \caption{Survival of spores in LISA simulator at  +23 and -80 $^\circ$C vs. exposure to UV light.}\label{fig:fig1}
\end{figure}

In our experiments, we found that desiccation  effect (water escape because of low pressure) may strongly decrease the survival of vegetative cells, but not of spores.
In Martian environment, UV light appears to have the most cytocidal effect, while atmospheric gases or temperature are not relevant to the survival of cells or spores. 
Vegetative cells are inactivated by UV light in a few minutes, while spores may survive for hours (Fig. \ref{fig:fig1}). Two of our {\it Bacillus} strains, {\it B. pumilus} and {\it B. Nealsonii}, have a particular capability to survive in Martian conditions without being screened by dust or other shields. As endospores suspension, they survive at least 4 hours and in some cases up to 28 hours in Martian conditions. 

We simulated the dust coverage happening on the real planet by blowing on the samples a very smaller quantity of grain of volcanic ash or dust of red iron oxide. Samples covered by these dust grains have shown a high percentage of survival, indicating that under the surface dust, if life was present on Mars in the past, some bacteria cell could still be present.

\begin{acknowledgments}
We would like to acknowledge the Air Liquide Italy - North East region for kindly providing us the liquid nitrogen as support of this research.  This work has been funded by the University of Padua funds (ex 60\%). 
\end{acknowledgments}

\end{document}